\documentclass[prx, 10pt, twocolumn, floatfix, groupedaddress, aps, longbibliography, nofootinbib]{revtex4-1}

\usepackage{times, mathrsfs, amsmath, amsfonts, graphics, graphicx, cancel, color, amsthm, bbm, mathtools, amssymb, physics}

\usepackage[nice]{nicefrac}
\usepackage[english]{babel}
\usepackage[margin=.75in]{geometry}
\usepackage[T1]{fontenc} 
\usepackage{tabularx,ragged2e,booktabs}
\usepackage{capt-of}
\newcolumntype{C}[1]{>{\Centering}m{#1}}

\usepackage{xfrac,relsize,float}
\usepackage{appendix}

\usepackage[unicode=true, breaklinks=false, pdfborder={0 0 1}, backref=false, colorlinks=true, linkcolor=blue, citecolor=blue]{hyperref}

\newcommand{\seq}{\,{=}\,}
\newcommand{\Id}{\mathbbm{1}}

\newcommand{\maxnorm}[1]{\|#1\|_{\mbox{max}}}

\DeclareMathOperator*{\argmin}{argmin}

\begin{document}
\title{Selective quantum state tomography}
\author{Joshua Morris}
\email{joshua.morris@univie.ac.at}
\affiliation{Vienna Center for Quantum Science and Technology (VCQ), Faculty of Physics, University of Vienna, Vienna, Austria}
\author{Borivoje Daki\'{c}}
\email{borivoje.dakic@univie.ac.at}
\affiliation{Vienna Center for Quantum Science and Technology (VCQ), Faculty of Physics, University of Vienna, Vienna, Austria}
\affiliation{Institute for Quantum Optics and Quantum Information (IQOQI), Austrian Academy of
Sciences, Vienna, Austria}
\date{\today}

\begin{abstract}
We introduce the concept of selective quantum state tomography or SQST, a tomographic scheme that enables a user to estimate arbitrary elements of an unknown quantum state using a fixed measurement record. We demonstrate how this may be done with the following notable advantages (i) a number of state copies that depends only on the desired precision of the estimation, rather than the dimension of the unknown state; (ii) a similar reduction in the requisite classical memory and computational cost; (iii) an approach to state tomography using $O(\epsilon^{-2}\log d)$ state copies for maximum norm error $\epsilon$, as well as achieving nearly optimal bounds for full tomography with independent measurements. As an immediate extension to this technique we proceed to show that SQST can be used to generate an universal data sample, of fixed and dimension independent size, from which one can extract the mean values from a continuous class of operators on demand. 
\end{abstract}

\maketitle

\section*{Introduction}

A significant hurdle facing the production of large scale quantum devices is maintaining control over the corresponding exponential growth in Hilbert space as additional subsystems are added. This growth is well understood\! \cite{sample_opt_haah,Wright_1,Wright_2} and a source of such devices advantage over their classical counterparts. Unfortunately, this very advantage introduces an intrinsic handicap; an exponentially large output will in general need to be probed through quantum state tomography\! \cite{QST_ariano}. Naturally as our ability to exploit quantum effects for real world applications expands\! \cite{shor_factor, QONN}, the severity of this scaling problem becomes more cumbersome. 

This is problematic as the number of quantum information tasks that require mapping an input to an unknown output is rather significant\! \cite{input_output}. Accurately determining this output state is of the utmost importance for complete utilisation of a quantum device but with a large system may be completely infeasible. Even with perfect, error free quantum operations it seems a significant obstacle of the near future is going to be having the answer but being unable to access it. While there are certainly ways to mitigate this effect -- perhaps by compressing meaningful system outputs to a much smaller and hence manageable state space -- these strategies can only ever be a stopgap. There comes a point where a quantum system is simply too large for us to meaningfully characterise or measure, marking a kind of dimension demarcation for any near-term quantum technology. 

Tomography schemes abound that aim to reduce the difficulty of this task. However, an unavoidable fact of estimating an arbitrary density operator is the required polynomial number of measurements in the dimension $d$ -- of order ${O}(d^{\alpha})$, for some constant $\alpha$. More precisely, achieving an absolute error $\epsilon$ in the estimation of an unknown density matrix requires at least ${O}(d^{2}\epsilon^{-2})$\! \cite{shadow_tom} copies of a quantum state. Moreover, these strategies require the use of a quantum computer and entangling gates which are still challenging to implement. For more practical strategies where a user is restricted to independent (between state copies) measurements the scaling becomes slightly worse, with ${O}(d^{3}\epsilon^{-2})$ copies required\! \cite{sample_opt_haah, low_rank_QST}. 

If one is willing to make certain assumptions about a target state, exceptional gains can be made\! \cite{MPS_tomog,eff_tomog, NN1, NN2, NN3}. For example, estimation of a rank $r$ density operator may be performed using ${O}(d r^2\epsilon^{-2})$\! \cite{Compressed_tom, compressed_Ivan} measurements, falling under the category of sparse estimation or compressed sampling. Though these kinds of strategies often differ drastically,
their commonality is the need for a number of measurements that is dependent on the dimension of the system being reconstructed -- an exponential scaling. 

If this was not enough, an often overlooked part of any tomography is the classical task of estimating an exponentially large complex matrix given some measurement record constructed from an experiment. Given that our ability to engineer classical computing systems far exceeds our ability to do the same in the quantum realm, the memory and computational power required in the post-processing phase of state tomography has been safely ignored. But as our ability to control quantum systems begins to match our skill in the classical regime, this computational cost becomes intractable, particularly in terms of the memory required to store a $d^2$ complex matrix and the computational power required to process it.

If, on the other hand, our goal is only to access particular information about the target state (as in shadow tomography\! \cite{shadow_tom} or direct fidelity estimation\! \cite{dir_flam}) then it would seem logical that we can do so without needing to store the entire density operator. In this paper we focus on the latter case, presenting a tomographic scheme, dubbed Selective Quantum State Tomography (SQST), that does just this; enabling the estimation of an arbitrary density operator element with a number of copies, memory and computational cost that scales as ${O}(\epsilon^{-2}\log \delta^{-1} )$ with error (of the operator element) $\epsilon$ and failure probability $\delta$. This scaling is entirely independent of the dimension of the system. Additionally, we also demonstrate that the scheme is nearly optimal for complete state tomography when restricted to independent measurements. 

Extending this, we provide a generalisation of SQST to the estimation of a mean value of an arbitrary operator. We demonstrate the existence of a continuous submanifold of bounded operators, from which the mean value of an arbitrary element can be estimated. This is at a constant cost using a fixed set of measurement outcomes whose cardinality is dimension independent. 


The procedure for SQST is straightforward and can be conceptually understood via Mutually Unbiased Bases
(MUB)\! \cite{Ivonovic_1981,wooters_mub_optimal}. Though a thorough review of this topic is encouraged, we shall restrict ourselves to only a brief summary of their most important properties. A more complete examination can be found in \! \cite{MUBs_Z}.

\section*{Mutually Unbiased Bases}

Mutually unbiased basis sets are groups of orthogonal bases defined on a finite dimensional (of dimension $d$) Hilbert space. They hold a special property whereby any two basis elements $\ket{i,m}$ and $\ket{j,n}$ drawn from different sets -- indexed as $m$ and $n$ -- have a constant inner product $|\ip{i,m}{j,n}|^2=1/d, \; \forall m\neq n$. Though the underlying theory behind MUBs is exceptionally deep, being heavily related to fundamental properties of quantum mechanics such as complementarity\! \cite{Bori_MUB} and quantum geometry\! \cite{MUBs_Z}, for us it suffices to know that in a $d=p^n$-dimensional Hilbert space (with $p$ prime and integer $n$) it is known that a maximal mutually unbiased basis set consists of $d+1$ orthogonal bases $\{\ket{k,m} \}$ with $k=1,\dots, d$ and $m=1,\dots, d+1$.

While there are infinitely many complete MUBs, we are always free to apply a global unitary to each element of the set, transforming them into a different one while maintaining the inner product between elements. Due to this, we will always choose the $m=1$ basis to be the computational basis and define the remaining bases in terms of this set
\begin{equation}\label{MUB_exp}
\ket{k,m} = \frac{1}{\sqrt{d}} \sum_{l=1}^{d-1}\alpha_{l}^{km}\ket{l,1}; \quad m\neq 1,
\end{equation}
with $|\alpha_l^{km}| = 1$. The specific form of $\alpha_l^{km}$ is dependent on the dimension of the underlying Hilbert space, with different expressions for prime\! \cite{Ivonovic_1981} and prime power\! \cite{Paterek_opdef} dimensions. 
For our pursuit of a selective tomography, we exploit a useful fact\! \cite{Paterek_opdef} about arbitrary operators $A$ acting on the same space our MUB is defined upon, namely that
\begin{equation}\label{op_rep}
A = - \tr(A)\Id + \sum_{m=1}^{d+1} \sum_{k=1}^{d} O_{k}^{(m)} \Pi_k^{(m)},
\end{equation}
with $O_{k}^{(m)} = \tr[A \cdot \Pi_k^{(m)} ]$. The $\Pi_k^{(m)}$ are constructed from the basis elements of the MUB such that $\Pi_k^{(m)} =  \outerproduct{k,m}$. Since our MUB is informationally complete, any operator $A$ can be decomposed in this way. This result will serve us well in our pursuit of a tomography scheme capable of extracting single elements of a density operator. A particularly critical example for us are the matrix unit operators. Let $A_{ij} \coloneqq \outerproduct{j}{i}$ with $\ket{i}$ defined in the computational basis and $i\neq j$. Since $\expval{A_{ij}} = \tr[\rho A_{ij}] = \rho_{ij}$, measuring a particular operator element $\rho_{ij}$ amounts to estimating the expectation value of $A_{ij}$. We note that $\tr[A_{ij}] = 0$ given $i\neq j$ and use Eq.\! \eqref{op_rep} to write
\begin{equation}
A_{ij} = \sum_{m=1}^{d+1} \sum_{k=1}^{d} \bra{k,m}A_{ij} \ket{k,m} \Pi_k^{(m)}.
\end{equation}
Inserting the MUB element representation from Eq.\! \eqref{MUB_exp}, we arrive at the result
\begin{equation}\label{eq:Aij}
A_{ij} =  \frac{1}{d}\sum_{m=2}^{d+1} \sum_{k=1}^{d} \eta_{ij}^{km} \Pi_k^{(m)},
\end{equation} 
with $ \eta_{ij}^{km} = \alpha_{i}^{km}\alpha_{j}^{km *}$ and $|\eta_{ij}^{km}|=1$, being confined to the complex unit circle. From Eq.\! \eqref{eq:Aij}, it appears that a particular $A_{ij}$ is a specific sum of weighted measurement projectors. Continuing this line of inquiry, let us construct a POVM with elements defined as $R_k^{(m)} = \Pi_k^{(m)}/d$ and thus 
\begin{equation}
    A_{ij} = \sum_{k,m} \eta_{ij}^{km} R_{k}^{(m)}.
\end{equation}
Computing the expectation value of $A_{ij}$ and setting the probability $p_{km} \seq \tr[\rho \cdot R_{k}^{(m)}]$ 
\begin{equation}\label{eq:exact_rhoij}
\rho_{ij} = \expval{A_{ij}}  = \sum_{km} \eta_{ij}^{km} p_{km},
\end{equation}
we can see that $\rho_{ij}$ amounts to the expectation value of a bounded random variable $\eta_{ij}^{km}$ associated to the POVM measurement outcomes. This value can be efficiently estimated in experiment.

\section*{Selective Tomography}

\begin{figure}[ht]
\includegraphics[width=0.45\textwidth]{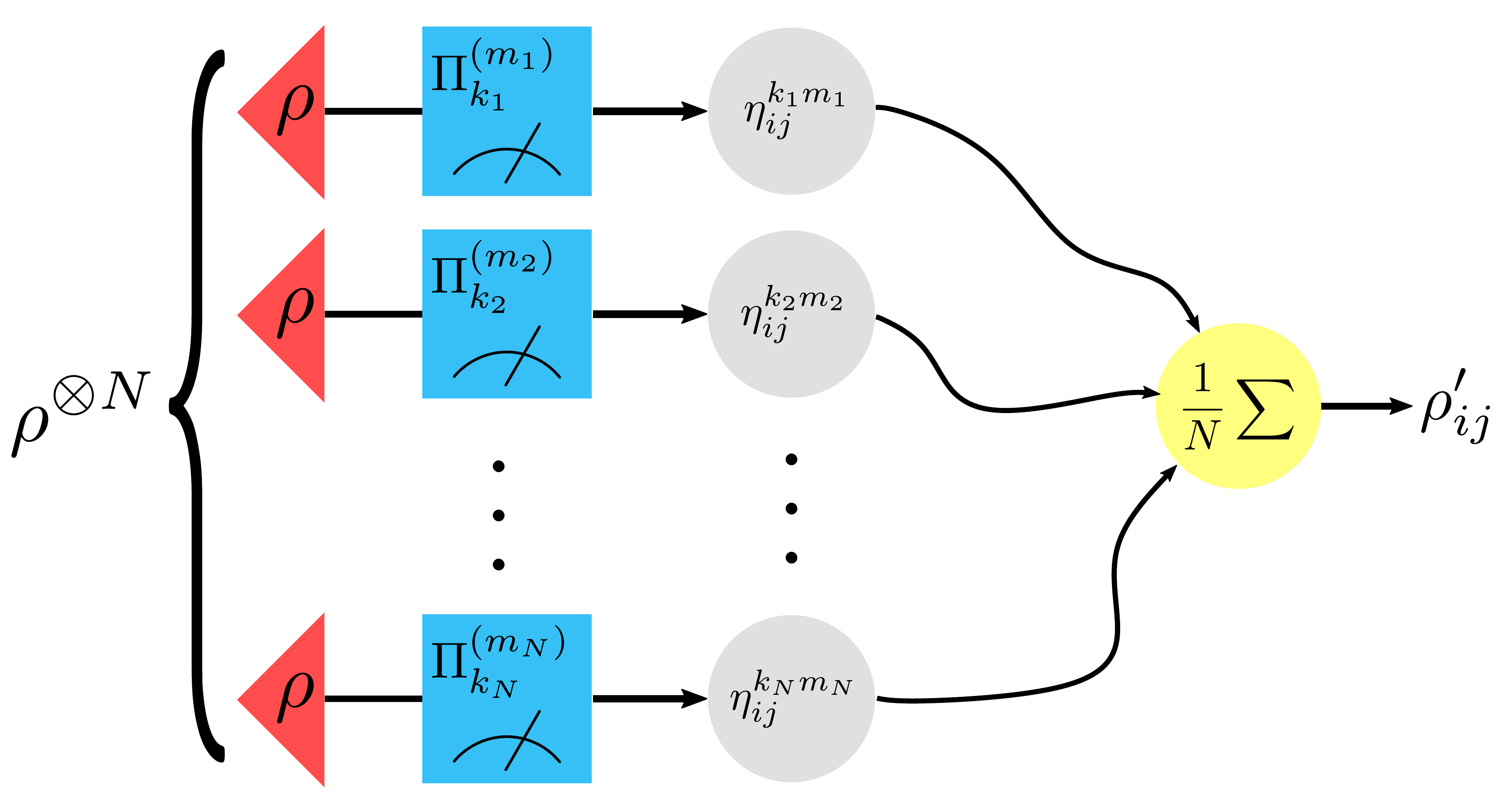}
\caption{The experimental procedure for estimation of a single element of an arbitrary density operator $\rho_{ij}$. $N$ copies of a state (red) are measured with a fixed POVM (blue), producing an outcome sequence. In post-processing, the $i,j$ are then fixed which, along with the outcome sequence, defines a term (grey) in the sum of Eq.\! \eqref{el_est}. The sum of these (yellow) form $\rho_{ij}^\prime$, the estimator for $\rho_{ij}$.}
\label{fig:experiment}
\end{figure}

In the previous section, we saw that the off-diagonal expectation values $\expval{A_{ij}}$ are equivalent to the expectation value of the random variable $\eta_{ij}^{(s)}  \in \{\eta_{ij}^{km} \,|\,m\seq 2 \dots d+1, k\seq \! 1 \dots d \}$, associated with outcomes of the POVM $\{R_k^{(m)}\}$ with $k,m$ indexed as in Eq.\! \eqref{eq:Aij}. 
Practical implementation of this POVM amounts to randomly choosing one of $d$ orthonormal basis sets (not including $m=1$) to measure a copy of $\rho$ in, each with probability $1/d$ of being selected. A tomography to estimate $\rho_{ij}$ would then proceed by the generation of $N$ copies of $\rho$, each measured using this POVM. For each measurement outcome, indexed by $s$, we update an approximation to the above sum as 
\begin{equation}\label{el_est}
    \rho_{ij}^\prime = \frac{1}{N}\sum_{s=1}^N \eta_{ij}^{(s)}.
\end{equation}
To be completely explicit, a selective quantum state tomography (Fig.\! \ref{fig:experiment}) would proceed in experiment as follows: 
\begin{enumerate}
    \item{Generate the quantum state $\rho$ to be measured, defined on a Hilbert space of dimension $d$.}
    \item{Measure this state using the POVM defined by $\{R_k^{(m)} \}$, obtaining measurement result $(k,m)$.} 
    \item{Using a total of $N$ copies of the state, repeat step two, generating the sequence of outcomes $\{(k_1,m_1),\dots (k_N,m_N) \}$. This concludes the experimental phase of the tomography.}
    \item{In post processing, compute the estimate of a particular $\rho_{ij}$ using indexes $i,j$ and $(k_s,m_s)$ to compute $\eta_{ij}^{(s)}$ and the sum in Eq.\! \eqref{el_est}.}
    \item{To estimate a different element $\rho_{ab}$, simply update the values of $i,j$ to $a,b$ and recompute the estimator, without further measurements.}
\end{enumerate}

With the experimental procedure now well-defined, our goal is to compute the number of state copies $N$ of $\rho$ required for the estimator $\rho_{ij}^\prime$ to converge to $\rho_{ij}$ within some error $\epsilon$ and failure probability $\delta$. Though $\eta_{ij}^{(s)}$ complicates matters by taking values on the complex unit circle, we may still apply the usual concentration inequalities by considering $\eta_{ij}^{(s)}$ as a sum of two bounded random variables such that $|\Re[\eta_{ij}^{(s)}]+i\Im[\eta_{ij}^{(s)}]| = 1$ . Recall that $\rho_{ij}^\prime = N^{-1}\sum_{s=1}^N \eta_{ij}^{(s)}$ and note that  $\mathbb{E}[\rho_{ij}^\prime] = \rho_{ij}$. Following a concentration inequality approach we wish to compute the bound $ \Pr(\left| \rho_{ij}^\prime - \mathbb{E}[\rho_{ij}^\prime] \right|\geq \epsilon)$. First, we will isolate the real and complex components of the random variable $\eta_{ij}^{(s)}$. By the triangle inequality we have that
\begin{equation*}
\begin{split}
\Pr(\left| \rho_{ij}^\prime - \mathbb{E}[\rho_{ij}^\prime] \right|\geq \epsilon) & \leq \Pr(\left| A \right| \geq \frac{\epsilon}{2} \cup \left| B \right|\geq \frac{\epsilon}{2}),\\
& \leq \Pr(|A|\geq \frac{\epsilon}{\sqrt{2}}) + \Pr(|B|\geq \frac{\epsilon}{\sqrt{2}}),
\end{split}
\end{equation*}
for $A \seq \Re(\rho_{ij}^\prime) - \mathbb{E}[\Re(\rho_{ij}^\prime)]$ and $B \seq \Im(\rho_{ij}^\prime) - \mathbb{E}[\Im(\rho_{ij}^\prime)]$. 
From here, we can apply Hoeffding's inequality\! \cite{hoeff} for bounded random variables to each term individually 
\begin{equation}\label{eq:POVM_bnd}
    \Pr(\left| \rho_{ij}^\prime - \mathbb{E}[\rho_{ij}^\prime] \right|\geq \epsilon) \leq 4e^{-\frac{N\epsilon^2}{2}} = \delta.
\end{equation}
From here we can infer the number of copies $N = {O}(\epsilon^{-2} \log \delta^{-1})$ needed to estimate $\rho_{ij}$ within the range $|\rho_{ij}^\prime - \rho_{ij}| < \epsilon$ with probability greater than $1-\delta$. This is in tandem with a ${O}(N)$ complexity overhead in both the required memory and computation, given we need only store the outcomes of each measurement and the summation may be computed piece-wise. 

\begin{figure}[ht]
\includegraphics[width=0.43\textwidth]{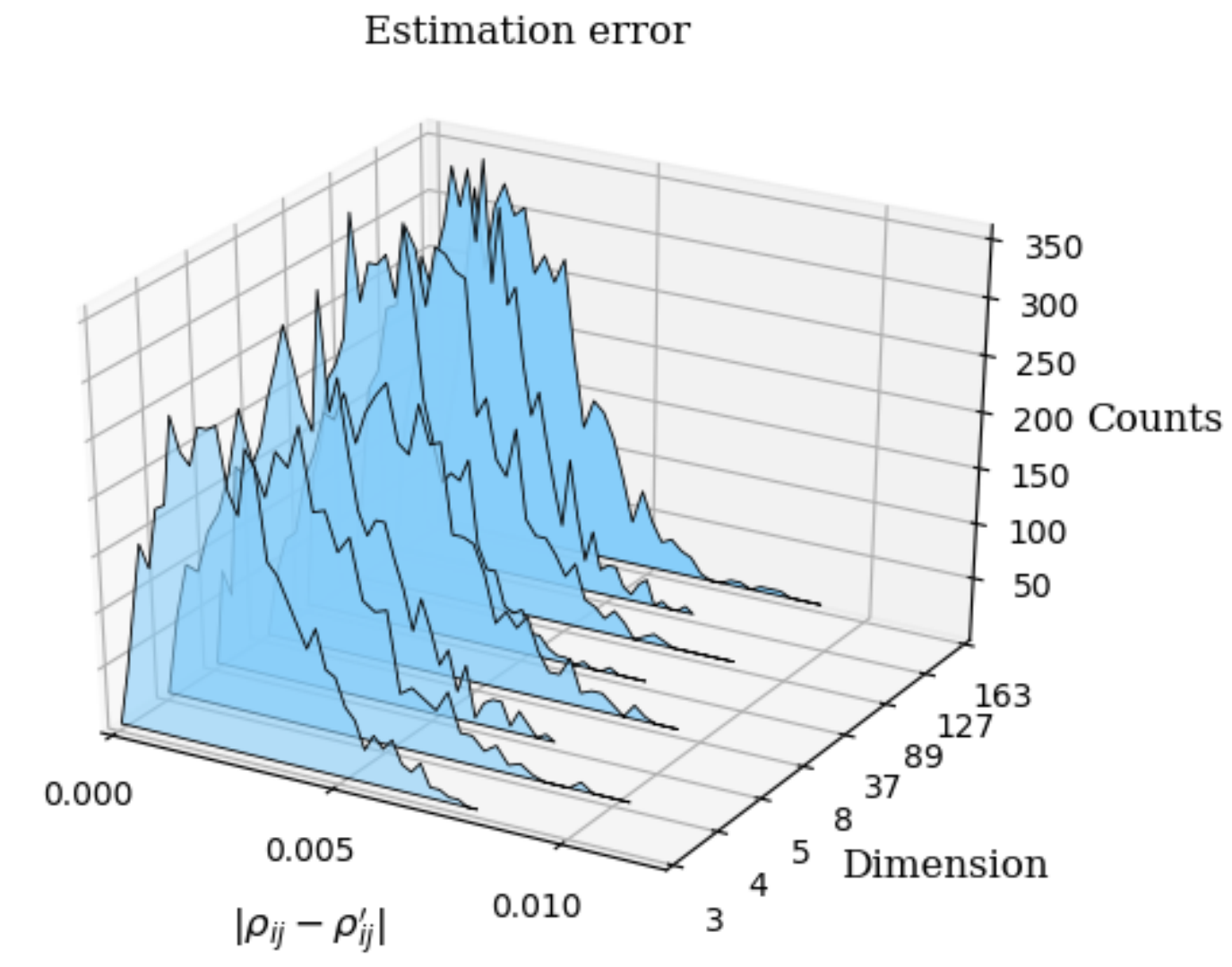}
\caption{Using approximately one hundred and twenty thousand copies each, one thousand superposition states $a\ket{i}+b\ket{j}$  are simulated and the off diagonal element $\outerproduct{i}{j}$ (with value $ab$) is estimated. By Eq. \ref{eq:copy_number}, we expect that an absolute error no greater than $10^{-2}$ will occur less than one percent of the time. Encouragingly, the error barely exceeds this bound even once with the three sigma deviation of each histogram falling well within our predicted bound.}
\label{fig:simulation}
\end{figure}

In pursuit of adding numerical evidence to support this claim, a challenging experimental task may be simulated using the SQST protocol on systems of a varying dimension $d$. Repeated one thousand times, a randomly chosen pair of computational basis states $\ket{i},\ket{j}$ with $i\neq j$ was chosen and a random superposition of them formed - see Figure \ref{fig:simulation}. For each instance, SQST is used to estimate the single off-diagonal of this state and bench-marked against the ground truth with the number of state copies going as in Eq. \ref{eq:POVM_bnd}. The results of this are summarised in Figure \ref{fig:simulation} and demonstrate the efficacy of the protocol. 

For estimation of any $\rho_{ij}$, we need also to account for the diagonal case $i=j$, something we neglect in the above formulation of SQST. Fortunately the estimation of the diagonal elements of $\rho_{ij}$ is straightforward. This stems from the fact that diagonal estimation of density operators is something of a simple case, achievable with measurement in the computational basis. For truly arbitrary estimation of the elements of a density operator we thus need to maintain two measurement records; one for the diagonal elements which gives the $\rho_{ii}$ directly, and another for the off diagonals $\rho_{ij}$, both requiring $N \seq {O}(\epsilon^{-2} \log \delta^{-1})$ copies of the state. An additional factor must be included if multiple elements are to be estimated, corresponding to $M$ repetitions of Step 5 in the experiment. For all of these estimated elements to \textit{simultaneously} have less than error $\epsilon$ with probability $1-\delta$, a union bound in Eq.\! \eqref{eq:POVM_bnd} is included for this more restrictive case. For $M$ estimated quantities this leads to a scaling of $N={O}(\epsilon^{-2} \log M \log \delta^{-1})$. As we shall soon show, this is the optimal scaling for this task.

\section*{Relation to Full State Tomography}

    Given that multiple elements may now be estimated simultaneously, it is natural to consider the extension of SQST to full state tomography. In this scenario we would estimate every element of the density operator ($M = d^2$) with a corresponding logarithmic cost. However the reconstructed 'state', so directly computed, has no guarantee of being a positive semi-definite matrix as required for a valid quantum state. The same problem plagues standard quantum state tomography with the usual solution being a maximum likelihood estimator that converges to the exact density operator in the asymptotic measurement limit\! \cite{Blume_Kohout_2010}. A similar estimator can be constructed here in the max norm and used to efficiently find the closest valid density operator by the same estimator projection argument detailed in\! \cite{murano, fast_state_kueng}. From Eq.\! \eqref{el_est} we have $\rho_{L} = \sum_{ij} \rho_{ij}'\outerproduct{i}{j}$ which by construction has max norm error no greater than $\epsilon$. 
Though it is not certainly positive, it is Hermitian and we can immediately compute the projection onto the set of positive semi-definite matrices 
\begin{equation}\label{eq::opt_est}
\rho_p \coloneqq \argmin_{\sigma\succeq 0} \maxnorm{\rho_{L} - \sigma},
\end{equation}
where $\maxnorm{A}=\mbox{max}_{ij}\,|A_{ij}|$. The optimisation problem itself may be phrased as a semi-definite program (SDP) 
\footnote{The problem of determining the density operator $Y$ that minimises the distance between it and an estimator $X$ in the max norm can be expressed as
\begin{equation*}
\begin{aligned}
& \underset{t,Y}{\text{minimize}}
& & t \\
& \text{subject to}
& & \left[ {\begin{array}{cc}
   t & X_{ij}-Y_{ij} \\
   X_{ij}^*-Y_{ij}^* & t \\
  \end{array} } \right]\succeq 0, \\
&&& Y \succeq 0.
\end{aligned}
\end{equation*}} that finds the density operator that minimises the max norm distance between it and the linear estimator $\rho_{L}$ which as an SDP can be computed efficiently\! \cite{boyd}. The failure probability of Eq.\! \eqref{eq::opt_est} can be bounded as $\Pr[\maxnorm{\rho_p- \rho}>\epsilon] \leq \Pr[\maxnorm{\rho_{L} - \rho}>\epsilon] \leq \delta$ with the right hand side bounds being the already computed case of $M=d^2$ elements. The proof follows from Appendix A of \cite{murano} and the fact that the triangle inequality holds equally well for the Frobenius and max norms. Given this, SQST allows for estimation of a complete density matrix with logarithmic measurement cost of $O(\epsilon^{-2} \log d \log \delta^{-1})$ and a bounded max norm error.

Naturally this result must come with a hefty caveat, as we now consider how this bounded error compares with a standard and more resource intensive full tomography in a more stringent norm. Such a comparison of our selective tomography with other tomographic schemes requires computing the bounds our scheme places on the norm of an error matrix $E = \rho - \rho^\prime$ where $\rho^\prime$ is the estimated density operator and $\rho$ is the ground truth. In standard quantum state tomography we require that the trace norm of this quantity is $||E||_1 \leq \nu$ for $\nu\geq 0$.
`
The estimation error of SQST is equivalent to the max norm $||E||_{\max}\coloneqq \max_{ij} |E_{ij}|\leq \epsilon$. This can be related to the trace distance norm via
\begin{equation}\label{eq:bounds}
    \frac{1}{\sqrt{d^3}}||E||_1 \leq ||E||_{\max} \leq ||E||_1,
\end{equation}
with the steps leading to these bounds may be found in the supplementary material. This inequality is informative in the following way: Suppose the presented scheme was used to estimate the value $\rho_{ij} $ of a state $\rho$ with an error known to be bounded by $||E||_{\max} = ||\rho_{ij} - \rho_{ij}^\prime||_{\max} \leq \epsilon$ using $N$ copies of the state. From Eq.\! \eqref{eq:bounds} this then implies an equivalent estimation error in the trace norm of the complete error matrix; $||E||_1 \leq  \sqrt{d^3}\epsilon = \nu$. We know that SQST requires ${O}(\epsilon^{-2} \log M \log \delta^{-1})$ copies for a maximum error of $\epsilon$ in $M=d^2$ elements, which combined with the previous bound gives a total of $N=\tilde{O}(d^3 \nu^{-2} \log \delta^{-1} )$\! \cite{polylog} state copies. This is provably optimal \! \cite{sample_opt_haah} with a $\log d$ overhead. This optimal scaling well justifies the choice of the max norm in Eq. \eqref{eq::opt_est} over the trace distance\! \cite{fast_state_kueng}, the standard used when considering problems of tomography.  
Notably, estimation in trace distance would only guarantee the same error scaling in the max norm, thus it does not provide optimality of SQST considered here. As shown, tomography in the max norm comes with a logarithmic cost while tomography in the trace distance will provide much worse scaling in max norm. This is to say that the bounds are unidirectional and one gains more information of the error by considering the max norm. 


\section*{Estimation of arbitrary operators}
So far we have confined ourselves to estimation of a discrete class of expectation values - the matrix unit operators formed from the computational basis. Here we present a generalisation of this technique to a continuous set of operators. Without loss of generality, let us focus our attention on the cases of traceless operators $\Tr A=0$. Consider a general decomposition given in Eq.\! \eqref{op_rep} of a traceless operator $A$
\begin{equation}
    A = \sum_{m=1}^{d+1}\sum_{k=1}^d a_{km}\Pi_{k}^{(m)}. 
 \end{equation}
Furthermore, we restrict our attention to bounded $|a_{km}|\leq K$. A straightforward calculation shows $||A||_2^2=\sum_{km}|a_{km}|^2\leq d(d+1)K^2$ (where $||\cdot ||_2$ is the Frobenius norm). In complete analogy to the SQST, the mean value of $A$ can be estimated by implementing the POVM $\{R_k^{(m)}=\Pi_{k}^{(m)}/(d+1)\}$ via an estimator as in Eq.\eqref{el_est}, i.e. $a^{\prime}=N^{-1}\sum_{s=1}^N a^{(s)}$, with $a^{(s)}\in\{(d+1)a_{km}|m=1\dots d+1,k=1\dots d\}$. Clearly $\mathbb{E}[a^\prime]=\mathbb{E}[A]$. With this the Hoeffding bound of Eq.\! \eqref{eq:POVM_bnd} now has a slightly modified form  
\begin{equation}\label{genA_bnd}
    \Pr(\left|a^\prime - \mathbb{E}[A] \right|\geq \epsilon) \leq 4e^{-\frac{N\epsilon^2}{2K^2(d+1)^2}} = \delta,
\end{equation}
therefore, the procedure requires 
\begin{equation}\label{eq:copy_number}
N = {O}(K^2d^2\epsilon^{-2} \log\delta^{-1}),
\end{equation}
state copies for reliable estimation. Now, as long as $K={O}(d^{-1})$, the number of required copies does not scale with dimension $d$.
The class of operators satisfying $|a_{km}|\leq K$ is a convex bounded set, with extreme points defined by $a_{km}=e^{i\varphi_{km}}/K$, i.e.
\begin{equation}
A_{\mathbf{\varphi}}=\frac{1}{K}\sum_{m=1}^{d+1}\sum_{k=1}^d e^{i\varphi_{km}}\Pi_{k}^{(m)}.  
\end{equation} 
By setting $K={O}(1/d)$, we have $||A_{\mathbf{\varphi}}||={O}(1)$, and the extreme points form a continuous sub-manifold of bounded operators parameterised by $d(d+1)$ real parameters $\varphi_{km}$. The mean value of any element $A_{\mathbf{\varphi}}$ can be extracted to the accuracy $\epsilon$ with the constant (size independent) overhead $N={O}(\epsilon^{-2} \log \delta^{-1})$. As before, if multiple operators $\{A_i\}_{i=1}^M$ need to be simultaneously estimated from the same set of measurements, a factor of $\log M$ must be included in the above.

\section*{Discussion}
We have presented a tomographic scheme for the estimation of a quantum state, capable of determining the value of individual density operator elements without an exponential number of measurements. This is performed in conjunction with no requirement to store or compute exponentially large matrices as intermediate steps. Since SQST can be applied to arbitrary density operators, it is conceivable that further reduction of the total number of measurements is achievable by combining it with system specific assumptions e.g. sparsity. Finally, while the POVM constructed from mutually unbiased bases may at first appear to be difficult to perform in experiment, it has been shown\! \cite{eff_MUB} that any MUB may be constructed using a universal gate set with linear cost. 


Combining these advantages of dimensional independence, ${O}(N)$ memory overhead, applicability to arbitrary quantum states and generality to a large class of bounded operators, we conclude that the presented tomography scheme is an experimentally advantageous approach to quantum state tomography. 

It is important to note that a similar method of direct extraction of density matrix elements is known for continuous variables (CV) systems as a method of measurement patterns\! \cite{D_Ariano_quadrature}. However, for CV systems it is know that the estimation error explicitly depends on the energy\! \cite{D_Ariano_quadrature,gheorg_veri}, i.e. the error for estimation of $\rho_{nm} $ worsens as $n$ and $m$ increase ($n,m$ are indexes of energy eigenstates). Curiously, we do not observe this in our SQST for discrete systems, with this fact forming an interesting point for future consideration.

\textit{Note - } In preparation of this manuscript, we became aware of another work\! \cite{classical_shadow} considering a similar problem. The authors of that work use a different measurement scheme exploiting random Clifford circuits that achieves the same sample complexity as the one presented here. 
\\
\\
\section*{acknowledgements}
We thank Michael Kewming for his thorough review of this paper and James Saunderson for his expertise on semidefinite programming. J.M and B.D both acknowledge support from the Austrian Science Fund (FWF) through BeyondC (F71).
\bibliography{SQST}

\begin{thebibliography}{32}%
\makeatletter
\providecommand \@ifxundefined [1]{%
 \@ifx{#1\undefined}
}%
\providecommand \@ifnum [1]{%
 \ifnum #1\expandafter \@firstoftwo
 \else \expandafter \@secondoftwo
 \fi
}%
\providecommand \@ifx [1]{%
 \ifx #1\expandafter \@firstoftwo
 \else \expandafter \@secondoftwo
 \fi
}%
\providecommand \natexlab [1]{#1}%
\providecommand \enquote  [1]{``#1''}%
\providecommand \bibnamefont  [1]{#1}%
\providecommand \bibfnamefont [1]{#1}%
\providecommand \citenamefont [1]{#1}%
\providecommand \href@noop [0]{\@secondoftwo}%
\providecommand \href [0]{\begingroup \@sanitize@url \@href}%
\providecommand \@href[1]{\@@startlink{#1}\@@href}%
\providecommand \@@href[1]{\endgroup#1\@@endlink}%
\providecommand \@sanitize@url [0]{\catcode `\\12\catcode `\$12\catcode
  `\&12\catcode `\#12\catcode `\^12\catcode `\_12\catcode `\%12\relax}%
\providecommand \@@startlink[1]{}%
\providecommand \@@endlink[0]{}%
\providecommand \url  [0]{\begingroup\@sanitize@url \@url }%
\providecommand \@url [1]{\endgroup\@href {#1}{\urlprefix }}%
\providecommand \urlprefix  [0]{URL }%
\providecommand \Eprint [0]{\href }%
\providecommand \doibase [0]{http://dx.doi.org/}%
\providecommand \selectlanguage [0]{\@gobble}%
\providecommand \bibinfo  [0]{\@secondoftwo}%
\providecommand \bibfield  [0]{\@secondoftwo}%
\providecommand \translation [1]{[#1]}%
\providecommand \BibitemOpen [0]{}%
\providecommand \bibitemStop [0]{}%
\providecommand \bibitemNoStop [0]{.\EOS\space}%
\providecommand \EOS [0]{\spacefactor3000\relax}%
\providecommand \BibitemShut  [1]{\csname bibitem#1\endcsname}%
\let\auto@bib@innerbib\@empty
\bibitem [{\citenamefont {{Haah}}\ \emph {et~al.}(2017)\citenamefont {{Haah}},
  \citenamefont {{Harrow}}, \citenamefont {{Ji}}, \citenamefont {{Wu}},\ and\
  \citenamefont {{Yu}}}]{sample_opt_haah}%
  \BibitemOpen
  \bibfield  {author} {\bibinfo {author} {\bibfnamefont {J.}~\bibnamefont
  {{Haah}}}, \bibinfo {author} {\bibfnamefont {A.~W.}\ \bibnamefont
  {{Harrow}}}, \bibinfo {author} {\bibfnamefont {Z.}~\bibnamefont {{Ji}}},
  \bibinfo {author} {\bibfnamefont {X.}~\bibnamefont {{Wu}}}, \ and\ \bibinfo
  {author} {\bibfnamefont {N.}~\bibnamefont {{Yu}}},\ }\bibfield  {title}
  {\enquote {\bibinfo {title} {Sample-optimal tomography of quantum states},}\
  }\href {\doibase 10.1109/TIT.2017.2719044} {\bibfield  {journal} {\bibinfo
  {journal} {IEEE Transactions on Information Theory}\ }\textbf {\bibinfo
  {volume} {63}},\ \bibinfo {pages} {5628--5641} (\bibinfo {year}
  {2017})}\BibitemShut {NoStop}%
\bibitem [{\citenamefont {O'Donnell}\ and\ \citenamefont
  {Wright}(2015)}]{Wright_1}%
  \BibitemOpen
  \bibfield  {author} {\bibinfo {author} {\bibfnamefont {R.}~\bibnamefont
  {O'Donnell}}\ and\ \bibinfo {author} {\bibfnamefont {J.}~\bibnamefont
  {Wright}},\ }\bibfield  {title} {\enquote {\bibinfo {title} {Efficient
  quantum tomography},}\ }\href {https://arxiv.org/abs/1508.01907} {\bibfield
  {journal} {\bibinfo  {journal} {arXiv:1508.01907}\ } (\bibinfo {year}
  {2015})}\BibitemShut {NoStop}%
\bibitem [{\citenamefont {O'Donnell}\ and\ \citenamefont
  {Wright}(2016)}]{Wright_2}%
  \BibitemOpen
  \bibfield  {author} {\bibinfo {author} {\bibfnamefont {R.}~\bibnamefont
  {O'Donnell}}\ and\ \bibinfo {author} {\bibfnamefont {J.}~\bibnamefont
  {Wright}},\ }\bibfield  {title} {\enquote {\bibinfo {title} {Efficient
  quantum tomography {II}},}\ }\href {https://arxiv.org/abs/1612.00034}
  {\bibfield  {journal} {\bibinfo  {journal} {arXiv:1612.00034}\ } (\bibinfo
  {year} {2016})}\BibitemShut {NoStop}%
\bibitem [{\citenamefont {D'Ariano}\ \emph {et~al.}(2003)\citenamefont
  {D'Ariano}, \citenamefont {Paris},\ and\ \citenamefont
  {Sacchi}}]{QST_ariano}%
  \BibitemOpen
  \bibfield  {author} {\bibinfo {author} {\bibfnamefont {G.~M.}\ \bibnamefont
  {D'Ariano}}, \bibinfo {author} {\bibfnamefont {M.~G.~A.}\ \bibnamefont
  {Paris}}, \ and\ \bibinfo {author} {\bibfnamefont {M.~F.}\ \bibnamefont
  {Sacchi}},\ }\bibfield  {title} {\enquote {\bibinfo {title} {Quantum
  tomography},}\ }\href {https://arxiv.org/abs/quant-ph/0302028} {\bibfield
  {journal} {\bibinfo  {journal} {arXiv:quant-ph/0302028}\ } (\bibinfo {year}
  {2003})}\BibitemShut {NoStop}%
\bibitem [{\citenamefont {Shor}(1997)}]{shor_factor}%
  \BibitemOpen
  \bibfield  {author} {\bibinfo {author} {\bibfnamefont {P.}~\bibnamefont
  {Shor}},\ }\bibfield  {title} {\enquote {\bibinfo {title} {Polynomial-time
  algorithms for prime factorization and discrete logarithms on a quantum
  computer},}\ }\href {\doibase 10.1137/S0097539795293172} {\bibfield
  {journal} {\bibinfo  {journal} {SIAM Journal on Computing}\ }\textbf
  {\bibinfo {volume} {26}},\ \bibinfo {pages} {1484--1509} (\bibinfo {year}
  {1997})}\BibitemShut {NoStop}%
\bibitem [{\citenamefont {Steinbrecher}\ \emph {et~al.}(2018)\citenamefont
  {Steinbrecher}, \citenamefont {Olson}, \citenamefont {Englund},\ and\
  \citenamefont {Carolan}}]{QONN}%
  \BibitemOpen
  \bibfield  {author} {\bibinfo {author} {\bibfnamefont {G.~R.}\ \bibnamefont
  {Steinbrecher}}, \bibinfo {author} {\bibfnamefont {J.~P.}\ \bibnamefont
  {Olson}}, \bibinfo {author} {\bibfnamefont {D.}~\bibnamefont {Englund}}, \
  and\ \bibinfo {author} {\bibfnamefont {J.}~\bibnamefont {Carolan}},\
  }\bibfield  {title} {\enquote {\bibinfo {title} {Quantum optical neural
  networks},}\ }\href {https://arxiv.org/abs/1808.10047} {\bibfield  {journal}
  {\bibinfo  {journal} {arXiv:1808.10047}\ } (\bibinfo {year}
  {2018})}\BibitemShut {NoStop}%
\bibitem [{inp()}]{input_output}%
  \BibitemOpen
  \href@noop {} {}\bibinfo {note} {{Arguably, all of them}}\BibitemShut
  {NoStop}%
\bibitem [{\citenamefont {Aaronson}(2017)}]{shadow_tom}%
  \BibitemOpen
  \bibfield  {author} {\bibinfo {author} {\bibfnamefont {S.}~\bibnamefont
  {Aaronson}},\ }\bibfield  {title} {\enquote {\bibinfo {title} {Shadow
  tomography of quantum states},}\ }\href {https://arxiv.org/abs/1711.01053}
  {\bibfield  {journal} {\bibinfo  {journal} {arXiv:1711.01053}\ } (\bibinfo
  {year} {2017})}\BibitemShut {NoStop}%
\bibitem [{\citenamefont {Kueng}\ \emph {et~al.}(2014)\citenamefont {Kueng},
  \citenamefont {Rauhut},\ and\ \citenamefont {Terstiege}}]{low_rank_QST}%
  \BibitemOpen
  \bibfield  {author} {\bibinfo {author} {\bibfnamefont {R.}~\bibnamefont
  {Kueng}}, \bibinfo {author} {\bibfnamefont {H.}~\bibnamefont {Rauhut}}, \
  and\ \bibinfo {author} {\bibfnamefont {U.}~\bibnamefont {Terstiege}},\
  }\bibfield  {title} {\enquote {\bibinfo {title} {Low rank matrix recovery
  from rank one measurements},}\ }\href {https://arxiv.org/abs/1410.6913}
  {\bibfield  {journal} {\bibinfo  {journal} {arXiv:1410.6913}\ } (\bibinfo
  {year} {2014})}\BibitemShut {NoStop}%
\bibitem [{\citenamefont {Lanyon}\ \emph {et~al.}(2017)\citenamefont {Lanyon},
  \citenamefont {Maier}, \citenamefont {Holz{\"a}pfel}, \citenamefont
  {Baumgratz}, \citenamefont {Hempel}, \citenamefont {Jurcevic}, \citenamefont
  {Dhand}, \citenamefont {Buyskikh}, \citenamefont {Daley}, \citenamefont
  {Cramer}, \citenamefont {Plenio}, \citenamefont {Blatt},\ and\ \citenamefont
  {Roos}}]{MPS_tomog}%
  \BibitemOpen
  \bibfield  {author} {\bibinfo {author} {\bibfnamefont {B.~P.}\ \bibnamefont
  {Lanyon}}, \bibinfo {author} {\bibfnamefont {C.}~\bibnamefont {Maier}},
  \bibinfo {author} {\bibfnamefont {M.}~\bibnamefont {Holz{\"a}pfel}}, \bibinfo
  {author} {\bibfnamefont {T.}~\bibnamefont {Baumgratz}}, \bibinfo {author}
  {\bibfnamefont {C.}~\bibnamefont {Hempel}}, \bibinfo {author} {\bibfnamefont
  {P.}~\bibnamefont {Jurcevic}}, \bibinfo {author} {\bibfnamefont
  {I.}~\bibnamefont {Dhand}}, \bibinfo {author} {\bibfnamefont {A.~S.}\
  \bibnamefont {Buyskikh}}, \bibinfo {author} {\bibfnamefont {A.~J.}\
  \bibnamefont {Daley}}, \bibinfo {author} {\bibfnamefont {M.}~\bibnamefont
  {Cramer}}, \bibinfo {author} {\bibfnamefont {M.~B.}\ \bibnamefont {Plenio}},
  \bibinfo {author} {\bibfnamefont {R.}~\bibnamefont {Blatt}}, \ and\ \bibinfo
  {author} {\bibfnamefont {C.~F.}\ \bibnamefont {Roos}},\ }\bibfield  {title}
  {\enquote {\bibinfo {title} {Efficient tomography of a quantum
  many-body~system},}\ }\href {\doibase 10.1038/nphys4244} {\bibfield
  {journal} {\bibinfo  {journal} {Nature Physics}\ }\textbf {\bibinfo {volume}
  {13}},\ \bibinfo {pages} {1158} (\bibinfo {year} {2017})}\BibitemShut
  {NoStop}%
\bibitem [{\citenamefont {Cramer}\ \emph {et~al.}(2010)\citenamefont {Cramer},
  \citenamefont {Plenio}, \citenamefont {Flammia}, \citenamefont {Somma},
  \citenamefont {Gross}, \citenamefont {Bartlett}, \citenamefont
  {{Landon-Cardinal}}, \citenamefont {Poulin},\ and\ \citenamefont
  {Liu}}]{eff_tomog}%
  \BibitemOpen
  \bibfield  {author} {\bibinfo {author} {\bibfnamefont {M.}~\bibnamefont
  {Cramer}}, \bibinfo {author} {\bibfnamefont {M.~B.}\ \bibnamefont {Plenio}},
  \bibinfo {author} {\bibfnamefont {S.~T.}\ \bibnamefont {Flammia}}, \bibinfo
  {author} {\bibfnamefont {R.}~\bibnamefont {Somma}}, \bibinfo {author}
  {\bibfnamefont {D.}~\bibnamefont {Gross}}, \bibinfo {author} {\bibfnamefont
  {S.~D.}\ \bibnamefont {Bartlett}}, \bibinfo {author} {\bibfnamefont
  {O.}~\bibnamefont {{Landon-Cardinal}}}, \bibinfo {author} {\bibfnamefont
  {D.}~\bibnamefont {Poulin}}, \ and\ \bibinfo {author} {\bibfnamefont
  {Y.}~\bibnamefont {Liu}},\ }\bibfield  {title} {\enquote {\bibinfo {title}
  {Efficient quantum state tomography},}\ }\href {\doibase 10.1038/ncomms1147}
  {\bibfield  {journal} {\bibinfo  {journal} {Nature Communications}\ }\textbf
  {\bibinfo {volume} {1}},\ \bibinfo {pages} {1--7} (\bibinfo {year}
  {2010})}\BibitemShut {NoStop}%
\bibitem [{\citenamefont {Torlai}\ \emph {et~al.}(2018)\citenamefont {Torlai},
  \citenamefont {Mazzola}, \citenamefont {Carrasquilla}, \citenamefont
  {Troyer}, \citenamefont {Melko},\ and\ \citenamefont {Carleo}}]{NN1}%
  \BibitemOpen
  \bibfield  {author} {\bibinfo {author} {\bibfnamefont {G.}~\bibnamefont
  {Torlai}}, \bibinfo {author} {\bibfnamefont {G.}~\bibnamefont {Mazzola}},
  \bibinfo {author} {\bibfnamefont {J.}~\bibnamefont {Carrasquilla}}, \bibinfo
  {author} {\bibfnamefont {M.}~\bibnamefont {Troyer}}, \bibinfo {author}
  {\bibfnamefont {R.}~\bibnamefont {Melko}}, \ and\ \bibinfo {author}
  {\bibfnamefont {G.}~\bibnamefont {Carleo}},\ }\bibfield  {title} {\enquote
  {\bibinfo {title} {Neural-network quantum state tomography},}\ }\href
  {\doibase 10.1038/s41567-018-0048-5} {\bibfield  {journal} {\bibinfo
  {journal} {Nature Physics}\ }\textbf {\bibinfo {volume} {14}},\ \bibinfo
  {pages} {447--450} (\bibinfo {year} {2018})}\BibitemShut {NoStop}%
\bibitem [{\citenamefont {Carrasquilla}\ \emph {et~al.}(2019)\citenamefont
  {Carrasquilla}, \citenamefont {Torlai}, \citenamefont {Melko},\ and\
  \citenamefont {Aolita}}]{NN2}%
  \BibitemOpen
  \bibfield  {author} {\bibinfo {author} {\bibfnamefont {J.}~\bibnamefont
  {Carrasquilla}}, \bibinfo {author} {\bibfnamefont {G.}~\bibnamefont
  {Torlai}}, \bibinfo {author} {\bibfnamefont {R.~G.}\ \bibnamefont {Melko}}, \
  and\ \bibinfo {author} {\bibfnamefont {L.}~\bibnamefont {Aolita}},\
  }\bibfield  {title} {\enquote {\bibinfo {title} {Reconstructing quantum
  states with generative models},}\ }\href {\doibase 10.1038/s42256-019-0028-1}
  {\bibfield  {journal} {\bibinfo  {journal} {Nature Machine Intelligence}\
  }\textbf {\bibinfo {volume} {1}},\ \bibinfo {pages} {155--161} (\bibinfo
  {year} {2019})}\BibitemShut {NoStop}%
\bibitem [{\citenamefont {Gao}\ and\ \citenamefont {Duan}(2017)}]{NN3}%
  \BibitemOpen
  \bibfield  {author} {\bibinfo {author} {\bibfnamefont {X.}~\bibnamefont
  {Gao}}\ and\ \bibinfo {author} {\bibfnamefont {L.}~\bibnamefont {Duan}},\
  }\bibfield  {title} {\enquote {\bibinfo {title} {Efficient representation of
  quantum many-body states with deep neural networks},}\ }\href {\doibase
  10.1038/s41467-017-00705-2} {\bibfield  {journal} {\bibinfo  {journal}
  {Nature Communications}\ }\textbf {\bibinfo {volume} {8}},\ \bibinfo {pages}
  {1--6} (\bibinfo {year} {2017})}\BibitemShut {NoStop}%
\bibitem [{\citenamefont {Gross}\ \emph {et~al.}(2010)\citenamefont {Gross},
  \citenamefont {Liu}, \citenamefont {Flammia}, \citenamefont {Becker},\ and\
  \citenamefont {Eisert}}]{Compressed_tom}%
  \BibitemOpen
  \bibfield  {author} {\bibinfo {author} {\bibfnamefont {D.}~\bibnamefont
  {Gross}}, \bibinfo {author} {\bibfnamefont {Y.}~\bibnamefont {Liu}}, \bibinfo
  {author} {\bibfnamefont {S.T.}\ \bibnamefont {Flammia}}, \bibinfo {author}
  {\bibfnamefont {S.}~\bibnamefont {Becker}}, \ and\ \bibinfo {author}
  {\bibfnamefont {J.}~\bibnamefont {Eisert}},\ }\bibfield  {title} {\enquote
  {\bibinfo {title} {Quantum state tomography via compressed sensing},}\ }\href
  {\doibase 10.1103/PhysRevLett.105.150401} {\bibfield  {journal} {\bibinfo
  {journal} {Phys. Rev. Lett.}\ }\textbf {\bibinfo {volume} {105}},\ \bibinfo
  {pages} {150401} (\bibinfo {year} {2010})}\BibitemShut {NoStop}%
\bibitem [{\citenamefont {Baldwin}\ \emph {et~al.}(2016)\citenamefont
  {Baldwin}, \citenamefont {Deutsch},\ and\ \citenamefont
  {Kalev}}]{compressed_Ivan}%
  \BibitemOpen
  \bibfield  {author} {\bibinfo {author} {\bibfnamefont {C.H.}\ \bibnamefont
  {Baldwin}}, \bibinfo {author} {\bibfnamefont {I.~H.}\ \bibnamefont
  {Deutsch}}, \ and\ \bibinfo {author} {\bibfnamefont {A.}~\bibnamefont
  {Kalev}},\ }\bibfield  {title} {\enquote {\bibinfo {title} {Strictly-complete
  measurements for bounded-rank quantum-state tomography},}\ }\href {\doibase
  10.1103/PhysRevA.93.052105} {\bibfield  {journal} {\bibinfo  {journal} {Phys.
  Rev. A}\ }\textbf {\bibinfo {volume} {93}},\ \bibinfo {pages} {052105}
  (\bibinfo {year} {2016})}\BibitemShut {NoStop}%
\bibitem [{\citenamefont {Flammia}\ and\ \citenamefont {Liu}(2011)}]{dir_flam}%
  \BibitemOpen
  \bibfield  {author} {\bibinfo {author} {\bibfnamefont {S.~T.}\ \bibnamefont
  {Flammia}}\ and\ \bibinfo {author} {\bibfnamefont {Y.}~\bibnamefont {Liu}},\
  }\bibfield  {title} {\enquote {\bibinfo {title} {Direct fidelity estimation
  from few pauli measurements},}\ }\href {\doibase
  10.1103/PhysRevLett.106.230501} {\bibfield  {journal} {\bibinfo  {journal}
  {Phys. Rev. Lett.}\ }\textbf {\bibinfo {volume} {106}},\ \bibinfo {pages}
  {230501} (\bibinfo {year} {2011})}\BibitemShut {NoStop}%
\bibitem [{\citenamefont {Ivonovi\'{c}}(1981)}]{Ivonovic_1981}%
  \BibitemOpen
  \bibfield  {author} {\bibinfo {author} {\bibfnamefont {I.~D.}\ \bibnamefont
  {Ivonovi\'{c}}},\ }\bibfield  {title} {\enquote {\bibinfo {title}
  {Geometrical description of quantal state determination},}\ }\href {\doibase
  10.1088/0305-4470/14/12/019} {\bibfield  {journal} {\bibinfo  {journal}
  {Journal of Physics A: Mathematical and General}\ }\textbf {\bibinfo {volume}
  {14}},\ \bibinfo {pages} {3241--3245} (\bibinfo {year} {1981})}\BibitemShut
  {NoStop}%
\bibitem [{\citenamefont {Wootters}\ and\ \citenamefont
  {Fields}(1989)}]{wooters_mub_optimal}%
  \BibitemOpen
  \bibfield  {author} {\bibinfo {author} {\bibfnamefont {W.~K.}\ \bibnamefont
  {Wootters}}\ and\ \bibinfo {author} {\bibfnamefont {B.~D.}\ \bibnamefont
  {Fields}},\ }\bibfield  {title} {\enquote {\bibinfo {title} {Optimal
  state-determination by mutually unbiased measurements},}\ }\href {\doibase
  https://doi.org/10.1016/0003-4916(89)90322-9} {\bibfield  {journal} {\bibinfo
   {journal} {Annals of Physics}\ }\textbf {\bibinfo {volume} {191}},\ \bibinfo
  {pages} {363 -- 381} (\bibinfo {year} {1989})}\BibitemShut {NoStop}%
\bibitem [{\citenamefont {Durt}\ \emph {et~al.}(2010)\citenamefont {Durt},
  \citenamefont {Englert}, \citenamefont {Bengsston},\ and\ \citenamefont
  {\'{Z}yczkowski}}]{MUBs_Z}%
  \BibitemOpen
  \bibfield  {author} {\bibinfo {author} {\bibfnamefont {T.}~\bibnamefont
  {Durt}}, \bibinfo {author} {\bibfnamefont {B.}~\bibnamefont {Englert}},
  \bibinfo {author} {\bibfnamefont {I.}~\bibnamefont {Bengsston}}, \ and\
  \bibinfo {author} {\bibfnamefont {K.}~\bibnamefont {\'{Z}yczkowski}},\
  }\bibfield  {title} {\enquote {\bibinfo {title} {On mutually unbiased
  bases},}\ }\href {\doibase 10.1142/S0219749910006502} {\bibfield  {journal}
  {\bibinfo  {journal} {International Journal of Quantum Information}\ }\textbf
  {\bibinfo {volume} {08}},\ \bibinfo {pages} {535--640} (\bibinfo {year}
  {2010})}\BibitemShut {NoStop}%
\bibitem [{\citenamefont {Paterek}\ \emph {et~al.}(2009)\citenamefont
  {Paterek}, \citenamefont {Daki\'{c}},\ and\ \citenamefont
  {Brukner}}]{Bori_MUB}%
  \BibitemOpen
  \bibfield  {author} {\bibinfo {author} {\bibfnamefont {T.}~\bibnamefont
  {Paterek}}, \bibinfo {author} {\bibfnamefont {B.}~\bibnamefont {Daki\'{c}}},
  \ and\ \bibinfo {author} {\bibfnamefont {C.}~\bibnamefont {Brukner}},\
  }\bibfield  {title} {\enquote {\bibinfo {title} {Mutually unbiased bases,
  orthogonal latin squares, and hidden-variable models},}\ }\href {\doibase
  10.1103/PhysRevA.79.012109} {\bibfield  {journal} {\bibinfo  {journal} {Phys.
  Rev. A}\ }\textbf {\bibinfo {volume} {79}},\ \bibinfo {pages} {012109}
  (\bibinfo {year} {2009})}\BibitemShut {NoStop}%
\bibitem [{\citenamefont {Wiesniak}\ \emph {et~al.}(2011)\citenamefont
  {Wiesniak}, \citenamefont {Paterek},\ and\ \citenamefont
  {Zeilinger}}]{Paterek_opdef}%
  \BibitemOpen
  \bibfield  {author} {\bibinfo {author} {\bibfnamefont {M.}~\bibnamefont
  {Wiesniak}}, \bibinfo {author} {\bibfnamefont {T.}~\bibnamefont {Paterek}}, \
  and\ \bibinfo {author} {\bibfnamefont {A.}~\bibnamefont {Zeilinger}},\
  }\bibfield  {title} {\enquote {\bibinfo {title} {Entanglement in mutually
  unbiased bases},}\ }\href
  {https://doi.org/10.1088%2F1367-2630%2F13%2F5%2F053047} {\bibfield  {journal}
  {\bibinfo  {journal} {New J. Phys. 13, 053047 (2011)}\ } (\bibinfo {year}
  {2011})}\BibitemShut {NoStop}%
\bibitem [{\citenamefont {Hoeffding}(1963)}]{hoeff}%
  \BibitemOpen
  \bibfield  {author} {\bibinfo {author} {\bibfnamefont {W.}~\bibnamefont
  {Hoeffding}},\ }\bibfield  {title} {\enquote {\bibinfo {title} {Probability
  inequalities for sums of bounded random variables},}\ }\href {\doibase
  10.1080/01621459.1963.10500830} {\bibfield  {journal} {\bibinfo  {journal}
  {Journal of the American Statistical Association}\ }\textbf {\bibinfo
  {volume} {58}},\ \bibinfo {pages} {13--30} (\bibinfo {year}
  {1963})}\BibitemShut {NoStop}%
\bibitem [{\citenamefont {Blume-Kohout}(2010)}]{Blume_Kohout_2010}%
  \BibitemOpen
  \bibfield  {author} {\bibinfo {author} {\bibfnamefont {Robin}\ \bibnamefont
  {Blume-Kohout}},\ }\bibfield  {title} {\enquote {\bibinfo {title} {Optimal,
  reliable estimation of quantum states},}\ }\href {\doibase
  10.1088/1367-2630/12/4/043034} {\bibfield  {journal} {\bibinfo  {journal}
  {New Journal of Physics}\ }\textbf {\bibinfo {volume} {12}},\ \bibinfo
  {pages} {043034} (\bibinfo {year} {2010})}\BibitemShut {NoStop}%
\bibitem [{\citenamefont {Sugiyama}\ \emph {et~al.}(2013)\citenamefont
  {Sugiyama}, \citenamefont {Turner},\ and\ \citenamefont {Murao}}]{murano}%
  \BibitemOpen
  \bibfield  {author} {\bibinfo {author} {\bibfnamefont {Takanori}\
  \bibnamefont {Sugiyama}}, \bibinfo {author} {\bibfnamefont {Peter~S.}\
  \bibnamefont {Turner}}, \ and\ \bibinfo {author} {\bibfnamefont {Mio}\
  \bibnamefont {Murao}},\ }\bibfield  {title} {\enquote {\bibinfo {title}
  {Precision-guaranteed quantum tomography},}\ }\href {\doibase
  10.1103/PhysRevLett.111.160406} {\bibfield  {journal} {\bibinfo  {journal}
  {Phys. Rev. Lett.}\ }\textbf {\bibinfo {volume} {111}},\ \bibinfo {pages}
  {160406} (\bibinfo {year} {2013})}\BibitemShut {NoStop}%
\bibitem [{\citenamefont {M.~Guta}(2018)}]{fast_state_kueng}%
  \BibitemOpen
  \bibfield  {author} {\bibinfo {author} {\bibfnamefont {R.~Kueng J. A.~Tropp}\
  \bibnamefont {M.~Guta}, \bibfnamefont {J.~Kahn}},\ }\bibfield  {title}
  {\enquote {\bibinfo {title} {1809.11162},}\ }\href
  {https://arxiv.org/abs/1809.11162} {\bibfield  {journal} {\bibinfo  {journal}
  {1809.11162}\ } (\bibinfo {year} {2018})}\BibitemShut {NoStop}%
\bibitem [{\citenamefont {Boyd}\ and\ \citenamefont
  {Vandenberghe}(2004)}]{boyd}%
  \BibitemOpen
  \bibfield  {author} {\bibinfo {author} {\bibfnamefont {Stephen}\ \bibnamefont
  {Boyd}}\ and\ \bibinfo {author} {\bibfnamefont {Lieven}\ \bibnamefont
  {Vandenberghe}},\ }\href@noop {} {\emph {\bibinfo {title} {Convex
  Optimization}}}\ (\bibinfo  {publisher} {Cambridge University Press},\
  \bibinfo {address} {USA},\ \bibinfo {year} {2004})\BibitemShut {NoStop}%
\bibitem [{pol()}]{polylog}%
  \BibitemOpen
  \href@noop {} {}\bibinfo {note} {{Following the convention of Aaronson (and
  others), $\tilde{O}$ denotes the complexity of $O$ with polylog factors
  suppressed.}}\BibitemShut {Stop}%
\bibitem [{\citenamefont {Seyfarth}\ and\ \citenamefont
  {Ranade}(2011)}]{eff_MUB}%
  \BibitemOpen
  \bibfield  {author} {\bibinfo {author} {\bibfnamefont {U.}~\bibnamefont
  {Seyfarth}}\ and\ \bibinfo {author} {\bibfnamefont {K.~S.}\ \bibnamefont
  {Ranade}},\ }\bibfield  {title} {\enquote {\bibinfo {title} {Construction of
  mutually unbiased bases with cyclic symmetry for qubit systems},}\ }\href
  {\doibase 10.1103/PhysRevA.84.042327} {\bibfield  {journal} {\bibinfo
  {journal} {Phys. Rev. A}\ }\textbf {\bibinfo {volume} {84}},\ \bibinfo
  {pages} {042327} (\bibinfo {year} {2011})}\BibitemShut {NoStop}%
\bibitem [{\citenamefont {D{\textquotesingle}Ariano}\ \emph
  {et~al.}(1996)\citenamefont {D{\textquotesingle}Ariano}, \citenamefont
  {Mancini}, \citenamefont {Man{\textquotesingle}ko},\ and\ \citenamefont
  {Tombesi}}]{D_Ariano_quadrature}%
  \BibitemOpen
  \bibfield  {author} {\bibinfo {author} {\bibfnamefont {G~M}\ \bibnamefont
  {D{\textquotesingle}Ariano}}, \bibinfo {author} {\bibfnamefont
  {S}~\bibnamefont {Mancini}}, \bibinfo {author} {\bibfnamefont {V~I}\
  \bibnamefont {Man{\textquotesingle}ko}}, \ and\ \bibinfo {author}
  {\bibfnamefont {P}~\bibnamefont {Tombesi}},\ }\bibfield  {title} {\enquote
  {\bibinfo {title} {Reconstructing the density operator by using generalized
  field quadratures},}\ }\href {\doibase 10.1088/1355-5111/8/5/007} {\bibfield
  {journal} {\bibinfo  {journal} {Quantum and Semiclassical Optics: Journal of
  the European Optical Society Part B}\ }\textbf {\bibinfo {volume} {8}},\
  \bibinfo {pages} {1017--1027} (\bibinfo {year} {1996})}\BibitemShut {NoStop}%
\bibitem [{\citenamefont {Gheorghiu}\ \emph {et~al.}(2018)\citenamefont
  {Gheorghiu}, \citenamefont {Kapourniotis},\ and\ \citenamefont
  {Kashefi}}]{gheorg_veri}%
  \BibitemOpen
  \bibfield  {author} {\bibinfo {author} {\bibfnamefont {Alexandru}\
  \bibnamefont {Gheorghiu}}, \bibinfo {author} {\bibfnamefont {Theodoros}\
  \bibnamefont {Kapourniotis}}, \ and\ \bibinfo {author} {\bibfnamefont
  {Elham}\ \bibnamefont {Kashefi}},\ }\bibfield  {title} {\enquote {\bibinfo
  {title} {Verification of quantum computation: An overview of existing
  approaches},}\ }\href {\doibase 10.1007/s00224-018-9872-3} {\bibfield
  {journal} {\bibinfo  {journal} {Theory of Computing Systems}\ }\textbf
  {\bibinfo {volume} {63}},\ \bibinfo {pages} {715--808} (\bibinfo {year}
  {2018})}\BibitemShut {NoStop}%
\bibitem [{\citenamefont {Huang}\ and\ \citenamefont
  {Kueng}(2019)}]{classical_shadow}%
  \BibitemOpen
  \bibfield  {author} {\bibinfo {author} {\bibfnamefont {H.}~\bibnamefont
  {Huang}}\ and\ \bibinfo {author} {\bibfnamefont {R.}~\bibnamefont {Kueng}},\
  }\bibfield  {title} {\enquote {\bibinfo {title} {Predicting features of
  quantum systems using classical shadows},}\ }\href
  {https://arxiv.org/abs/1908.08909} {\bibfield  {journal} {\bibinfo  {journal}
  {arXiv:1908.08909}\ } (\bibinfo {year} {2019})}\BibitemShut {NoStop}%
\end{thebibliography}%
\end{document}


\maketitle
\section{Error Bounds}\label{app:a}
In order to compare how an SQST tomography compares to a complete tomography, we consider how a bounded error in the max norm compares to the Frobenius norm as
\begin{equation}\label{max_upper}
    ||E||_{\max} = \max_{ij} |E_{ij}| \leq \sqrt{\sum_{ij} |E_{ij}|^2} = ||E||_2,
\end{equation}
while as a lower bound we have
\begin{equation}\label{max_lower}
    ||E||_2^2 = \sum_{ij}|E_{ij}|^2 \leq d^2 |E_{\max}|^2 = d^2 ||E||_{\max}^2,
\end{equation}
and hence $ ||E||_2 \leq d ||E||_{\max}$. The Frobenius norm $||\cdot ||_2$ is  a special case ($p=2$) of the Schatten $p$-norms whereas the trace norm corresponds to the  $p=1$ case. By monotonicity of the Schatten norm\! \cite{Bhatia1997} in $p$ we have that $||E||_q \leq ||E||_p$ for $ p<q$ and by H\"{o}lders inequality\!  \cite{Holder}
\begin{equation}
    ||AB||_1 \leq ||A||_p ||B||_q,\quad \forall\; \frac{1}{p}+\frac{1}{q} = 1,
\end{equation}
with $A=E,B=\Id, p=q=2$ we have that $||E||_1 \leq \sqrt{d} ||E||_2$. Combining this with Eqs.\! \eqref{max_upper},\eqref{max_lower} gives an inequality relation for the trace norm and maximum element wise error for SQST
\begin{equation}\label{eq:bounds}
    \frac{1}{\sqrt{d^3}}||E||_1 \leq ||E||_{\max} \leq ||E||_1.
\end{equation}